\def\la{\;
\raise0.3ex\hbox{$<$\kern-0.75em\raise-1.1ex\hbox{$\sim$}}\; }
\def\ga{\;
\raise0.3ex\hbox{$>$\kern-0.75em\raise-1.1ex\hbox{$\sim$}}\; }
\documentstyle[12pt,moriond,psfig]{article}
\begin{document}
\heading{ON THE MEASUREMENT OF KINETIC TEMPERATURE  
IN HIGH REDSHIFT GALACTIC HALOS}
\author{Sergei A. Levshakov $^{1}$,  Wilhelm H. Kegel $^{2}$} 
{$^{1}$ Department of Theoretical Astrophysics, Ioffe Institute, 
St.Petersburg, Russia.} 
{$^{2}$ Institut f\"ur Theoretische Physik der 
Universit\"at Frankfurt am Main, Frankfurt/Main, Germany.}

\begin{moriondabstract}
We consider the deconvolution of the thermal and the turbulent
profiles for a pair of ions with different atomic weights under
the condition that the lines are formed in an absorbing medium
with fluctuating density and velocity fields.
Our method is based on the entropy-regularized 
$\chi^2$-minimization
(ERM) procedure. From synthetic spectra we found that the CII and
SiII lines can be used to estimate the mean (density weighted)
kinetic temperature with a sufficiently high accuracy ($\sim $ 10\%) 
in a rather wide range of the ionization parameter,
$-6.0 \la \log U \la -2.0$.
This is a generalization of a previous result \cite{lta} where
we did not account for density fluctuations.

\end{moriondabstract}

\section{Introduction}

One of the main source of information about the 
physical properties of the intergalactic gas
is the study of metal-line absorption in quasar spectra.
Metal lines in the Lyman Limit Systems (LLS), -- 
i.e. in systems with $N_{\rm HI} \la 2\times10^{17}$~cm$^{-2}$
which are optically thin in the Lyman continuum, -- are of particular
interest in this regard because the Lyman continuum opacity does
not affect the metal abundance measurements in this case
\cite{ds}. Besides the LLSs often show carbon and
silicon line absorption from different ionization states which 
allows estimating the ionization parameter $U$ 
(the ratio of the number of photons with energies above one 
Rydberg to the number of atoms, $U = n_\gamma/n_{\rm H}$).   

The LLSs are usually assumed to arise in the outer regions of
intervening galactic halos where the electron density is rather
low, $n_e \sim 10^{-2} - 10^{-3}$~cm$^{-3}$. For such tenuous gas
the collisional ionization is not important in determining the
ionization fractions of ions since typical kinetic temperatures 
for the gas showing absorption in CII~--~CIV lines and in SiII~--~SiIV lines 
are of the order $10^4 - 2\times10^4$~K. 
It follows, that the outer parts of galactic halos are mainly
photoionized and that
the thermal and the ionization state of the gas may be
specified by the ionization parameter $U$ only,
\cite{ds}. Thus, for a given value of $n_\gamma$ (or the specific 
radiation flux $J_0$ at 1~Rydberg), the dispersion of $U$ along the 
line of sight represents a varying gas density.

In recent years, great efforts have been made towards 
the precise measurements of metal line profiles in QSO spectra
obtained with high spectral resolution, FWHM $\sim 5-7$~km~s$^{-1}$.
However, current theoretical models cannot give a complete prediction
to match the observational data \cite{kt}.

The observations often show complex structures of the line profiles.
It is traditional to treat them using the standard Voigt fitting
deconvolution procedure. 
This procedure is based on the assumption that
the apparent fluctuations within the line profile are caused by 
density clumps (`cloudlets') with different radial velocities.
The non-thermal (turbulent) velocity field inside
each cloudlet is accounted for in the so-called {\it microturbulent}
approximation. It was shown in \cite{lkm}, however, that the microturbulent
analysis may produce unphysical kinetic temperatures.

The more general 
{\it mesoturbulent} approach (see \cite{lkt} and the references
cited therein) is based on the
assumption that the intensity fluctuations
within the line profile arise mainly
from the irregular Doppler shifts in the absorption
coefficient caused by macroscopic large-scale, rather than thermal,
motions. 
If the macroscopic velocity field has a correlation length not small as
compared with the linear size of the absorbing region, then
the radial velocity distribution
may deviated significantly from the Gaussian model.

In a previous paper \cite{lta} we developed a method 
aimed at recovering the kinetic temperature from complex metal-line
spectra assuming a homogeneous gas density and a random velocity field.
Here we outline our first results for the case when both the density and
the velocity fields are of random nature. 

\section{The ERM procedure and results}

The  entropy-regularized $\chi^2$-minimization (ERM) procedure
utilizes complex but similar absorption line profiles of
ions with different atomic weights to estimate the mean
value of $T_{\rm kin}$ for the whole absorbing
region. The similarity of the complex profiles of ions with different masses
and ionization potentials stems from 
the equal ionization fraction for both of them. To illustrate this statement
we used a model thoroughly described in \cite{ds}~: an optically thin gas
ionized by a typical QSO photoionizing spectrum \cite{mf} with the
metallicity ${\rm Z}/{\rm Z}_\odot = 0.1$.

The consequent steps of our computational experiment are shown in Fig.~1.
Panel (a) presents the random velocity field $u(x) = v(x)/\sigma_{\rm turb}$,
where $x$ is the space coordinate 
$s$ along the line of sight in units of the linear
size $L$ of the absorbing region. The fluctuating density field $n(x)/n_0$
is depicted in panel (b). Both the velocity and the gas density fields were
calculated using the moving average method described in \cite{lta}.
To obtain the real gas density fluctuations, we used the log-normal 
distribution for the density contrast $\delta = (n - n_0)/n_0$
with the rms value of $\sigma_\delta = 1$ and $n_0 = 2\times10^{-3}$~cm$^{-3}$.
The rms value for the velocity field was assumed to be 20~km~s$^{-1}$, and
the linear size of the region $L = 5$~kpc.  
The chosen specific radiation flux 
$J_0 = 5\times10^{-22}$~ergs cm$^{-2}$ s$^{-1}$ Hz$^{-1}$ corresponds
in our case to $U_0 = 1.3\times10^{-3}$. 
In panel (c) we plot the equilibrium
kinetic temperature $T_4(x)$ in units $10^4$~K as a function of the ionization
parameter $U(x)$ in accord with the numerical results \cite{ds}. 
The fluctuations of $U(x)$ are shown in panel (d).
For each species we calculated the density weighted mean
temperature
\begin{equation}
\langle T \rangle = \frac{\int n_{\rm ion}(s)\,T(s)\,ds}
{\int n_{\rm ion}(s)\,ds}\: .
\label{eq:E1}
\end{equation}
The corresponding values of $\langle T \rangle$ we find for
CII, SiII, CIV, and SiIV are 15120~K, 15140~K, 17120~K, and 16390~K,
respectively.

These random fields lead in turn to complex profiles of 
CII$\lambda1334$, SiII$\lambda1260$, CIV$\lambda1548$, and SiIV$\lambda1394$. 
All profiles have been convolved with a Gaussian
spectrograph function having the width of 7~km~s$^{-1}$ (FWHM),
and then a Gaussian noise of S/N = 75 has been added.
The final patterns are shown
in panels (e), (f), (g), and (h), respectively, by dots
with corresponding error bars. 

It is seen that the CII and SiII profiles are similar whereas the 
intensity fluctuations within the CIV profile differ from those for
the SiIV line. This different behavior of low and high ionized species
formed in the same absorbing region is clearly illustrated in panel (i).
The ratio of fractional ionization for the CII and SiII lines is almost
constant over the whole range of $U$-values, but the corresponding ratio for the
CIV and SiIV lines is highly sensitive to the variation of $U$.

Next we considered these spectra as `observed' and analyzed them in two
ways. At first we applied the standard Voigt profile fitting procedure.
The results are shown in Figures 1e--1h, in where the individual components
are indicated by tick marks and the numbers give the corresponding
parameters (the first line gives the column density, the second the $b$
parameter. In Fig.~1h the third line gives the derived temperature). The
smooth curves show the resulting `theoretical' profiles. We see that
the derived temperatures vary between 2800~K and 45900~K whereas the
input temperature (Fig.~1c) varies only between 14360~K and 22090~K.
Contrary, when we apply our new ERM procedure \cite{lta} to the
CII and SiII lines we find from Fig.~1j (shaded area) a temperature of
$15200 \pm 200$~K. This value is in good agreement with the density
averaged temperatures derived before.

This result reinforces our previous finding that the Voigt profile
fitting procedure may lead to unphysical temperatures. On the other
side it indicates that -- at least for the cases studied so far --
our new procedure gives a physically reasonable average value
of the kinetic temperature.\\

This is a report on work in progress.\\

{\it Acknowledgment}. This work has been supported by the
Deutsche Forschungsgemeinschaft. SAL thanks the conference
organizers for financial assistance.

\vfill

\begin{figure*}
\vspace{-2.0cm}
\hspace{0cm}\psfig{figure=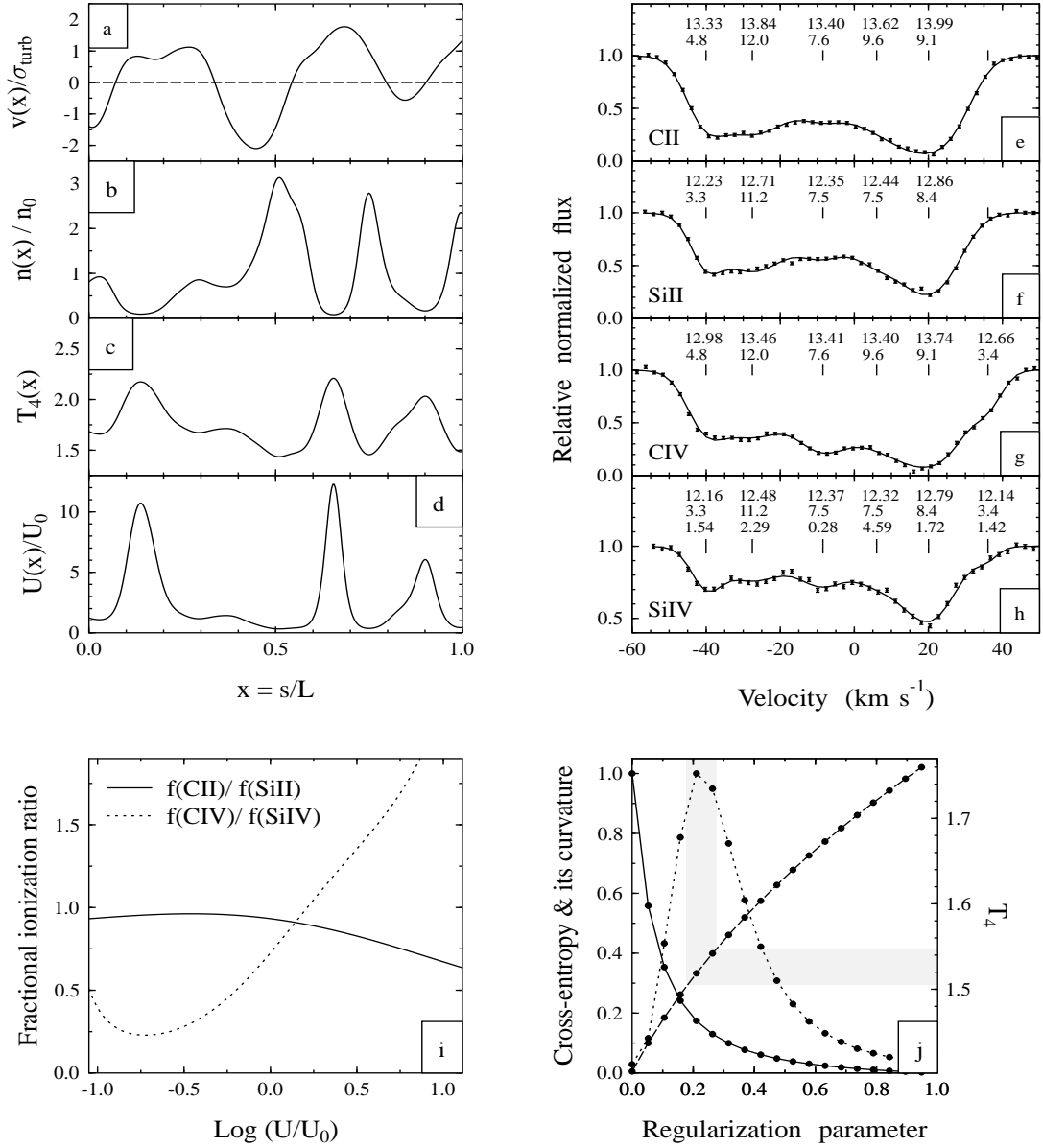,height=21.0cm,width=17.0cm}
\vspace{-3.7cm}
\caption[]{An example of Monte-Carlo simulations. ($a$) One realization
of the stochastic velocity field 
($\sigma_{\rm turb} = 20$~km~s$^{-1}$), ($b$) the density field  
($n_0 = 0.002$ cm$^{-3}$), ($c$) the kinetic temperature field (in
units $10^4$~K), and ($d$) the ionization parameter field 
($U_0 = 1.3\times10^{-3}$) versus the normalized space coordinate. 
($e,f,g,h$) Velocity plots of the simulated spectra (dots and
$1\sigma$ error bars) and the Voigt profile fits to the CII, SiII,
CIV, and SiIV lines (smooth curves). The column density and velocity
dispersion are displayed on top of each line. The third line in panel
($h$) shows the derived temperature for each subcomponent.
($i$) Fractional ionization ratios for the pair CII/SiII and CIV/SiIV
lines versus the normalized ionization parameter. ($j$) 
The normalized cross-entropy (filled circles connected by smooth line),
its curvature (dotted line), and
the kinetic temperature (dashed line) versus the normalized 
regularization parameter $\hat{\alpha}$ (for more details see 
\cite{lta}). The shaded area restricts the region of optimal values
of $\hat{\alpha}$ and corresponding solutions for $T_{\rm kin}$.  
}
\end{figure*}

\end{document}